\documentclass[12pt,preprint]{aastex}

\slugcomment{Submitted to AJ}
\shortauthors{Howell et al.}
\shorttitle{CVs in the {\it Kepler} Field}

\begin{document}

\title{
Spectroscopy of New and Poorly Known Cataclysmic Variables in the {\it Kepler} Field
}

\author{
Steve B. Howell  \altaffilmark{1,10,11}
Mark E. Everett  \altaffilmark{2,10}
Sally A. Seebode \altaffilmark{3,11}
Paula Szkody     \altaffilmark{4,10}
Martin Still     \altaffilmark{1,5}
Matt Wood     \altaffilmark{6}
Gavin Ramsay     \altaffilmark{7}
John Cannizzo     \altaffilmark{8}
Alan Smale     \altaffilmark{9}
}

\altaffiltext{1}{NASA Ames Research Center, Moffett Field, CA 94035}
\altaffiltext{2}{National Optical Astronomy Observatory, 950 N. Cherry Ave, 
Tucson, AZ 85719}
\altaffiltext{3}{San Mateo High School, San Mateo, CA 94401}
\altaffiltext{4}{Astronomy Department, University of Washington, Seattle, WA, 98195}
\altaffiltext{5}{Bay Area Environmental Research Inst, Inc., 560 Third St., West Sonoma, CA 95476}
\altaffiltext{6}{Physics \& Astronomy Dept., Texas A\&M University-Commerce, Commerce, TX, 75429}
\altaffiltext{7}{Armagh Observatory, College Hill, Armagh BT61 9DG, UK}
\altaffiltext{8}{CRESST and Astroparticle Physics Laboratory, NASA/GSFC, Greenbelt, MD, 20771 and Dept. of Physics, University of
Maryland, Baltimore County, 1000 Hilltop Circle, Baltimore, MD 21250}
\altaffiltext{9}{NASA Goddard Space Flight Center, Greenbelt, MD, 20771}
\altaffiltext{10}{Visiting Astronomer, Kitt Peak National Observatory}
\altaffiltext{11}{Visiting Astronomer, Mt. Palomar Observatory}


\begin{abstract} 
The NASA {\it Kepler} mission has been in science operation since May 2009 and is 
providing high precision, high cadence light curves
of over 150,000 targets. Prior to launch, nine cataclysmic variables were known to lie within {\it Kepler's} field of
view. We present spectroscopy for seven systems, four of which were
newly discovered since launch. 
All of the stars presented herein have been observed by, or are currently being observed by, 
the {\it Kepler} space telescope.
Three historic systems and one new candidate could not be detected at their
sky position and two candidates 
are called into question as to their true identity. 
\end{abstract}

\section{Introduction}

Cataclysmic Variables (CVs) 
are a well known group of interacting binary stars consisting of a white dwarf primary
and a low-mass companion. The two stars are in a tight orbit having binary periods 
from about 60 minutes to 8 hr, and
as such material flows from the low mass stars inner Lagrangian point toward the white dwarf.
This material, the accretion stream, has orbital angular momentum and as such forms an accretion disk
around the primary. Material within the disk is believed to
pile up and reach a critical density, causing an increase in temperature and eventual hydrogen ionization. 
The resulting disk brightening, an outburst, lasts a few days to a week or so and increases the CV's
light by a few to $\sim$4-6 magnitudes. Short period CVs, having periods less
than about 3 hours, show occasional larger (up to 9 magnitudes) 
and longer (up to a month) eruptions termed superoutbursts. These systems comprise the subgroup of CVs called the 
SU UMa stars. CVs with orbital periods approximately in the 4-6 hour range generally 
have the highest mass transfer rates (see Howell et al.,
2001) and show no outbursts. These binaries are believed to be in a state of 
constant outburst and are called novalike variables. Warner (2003) provides a detailed description
of all the types of cataclysmic variables.

Photometric observations of CVs reveal not only the semi-periodic outbursts but a wide variety of 
other phenomena related to mass transfer and accretion processes. Understanding the physics behind the behavior of 
the accretion disk, the outbursts
and the mass accretion process itself is often hampered by a lack of accurate or even well constrained
values for the basic stellar properties such as mass and temperature. 
The NASA {\it Kepler} mission (Borucki et al. 2010) 
is providing nearly continuous photometric coverage of the stars discussed
here and
spectroscopic work, in particular phase-resolved spectroscopy, will
yield good estimates for the stellar and binary properties. The {\it Kepler} light curves
consist of long cadence (30 min) and short
cadence (1 min) photometry and are all publicly available at the MAST archive\footnote{http://archive.stsci.edu/} 
and at the NASA exoplanet archive\footnote{http://exoplanetarchive.ipac.caltech.edu/}.

Scaringi et al. (2012) have reported on their initial discovery of 
11 new CV candidates within the
{\it Kepler} field of view based on blue + H$\alpha$ color selection. 
None of their sources are the same as those included in this paper and, at present,
none of their stars have {\it Kepler} observations.
Additional cataclysmic variables in the {\it Kepler} field, but not discussed in this paper, 
have had their {\it Kepler} light curves presented in
Barclay et al. (2012), Fontaine et al. (2011), Ramsay et al. (2012), Ostensen et al., (2010),
and Williams et al. (2010).

We report here on eleven CVs in the {\it Kepler}
field for which light curves are in hand (public) or forthcoming. 
Some of the CVs are, at best, poorly studied, four new CV candidates are discussed, and 
two systems may not 
be CVs at all. 

\section{Observations}

Nine cataclysmic variables were known to reside in in the {\it Kepler} field prior to the launch of this space
telescope in March 2009. 
Table 1 lists these nine historic systems as well as four new candidates. Eleven of these systems are discussed herein.
We do not discuss MV Lyr (Linnel et al. 2005) or V447 Lyr (Ramsay et al. 2012) in this paper.
The historic systems are catalogued in Downes and Shara (1993), 
Downes, Webbink, and Shara (1997) and in the on-line version of their catalog 
\footnote{http://archive.stsci.edu/prepds/cvcat/}. These catalogues will be referred to as
DSW97.
Our spectroscopic observations were obtained using the Kitt
Peak National Observatory 4-m telescope and the Mount Palomar 200" telescope. 
We describe the instrumental setups used for our spectroscopy at the two telescopes below.
Table 2 provides our spectroscopic observing log.
{\it Kepler} light curves used herein are the public, standard products and were 
obtained from the {\it Kepler} data archive at MAST. 

\subsection{Mount Palomar 200" Hale Telescope}

At Mt. Palomar, we used the double beam spectrograph (DBSP) attached to the 200" telescope. The dichroic filter
D-55 was used to split light between the blue and red arms. The blue arm used a 1200 l/mm grating providing
R$\sim$7700 and covered 1500\AA\ of spectrum. The red arm used a 1200 l/mm grating providing R$\sim$10000 and
covered only 670\AA. The slit width was set to 1 arcsec and the usual procedures of observing spectrophotometric stars
and arc lamps were adhered to. Red spectra were wavelength calibrated with a HeNeAr lamp while the blue arm
used a FeAr lamp. The nights were clear and provided stable seeing near 1 arcsec.
Data reduction was done using IRAF 2-D and 1-D routines for spectroscopic data and produced
a final 1-D spectrum for each observation. 
The Palomar instrument has a CCD acquisition camera which can be used
to take timed exposures of the local field of view as well as help to find and place faint sources 
on the slit. Integrations with this camera allowed us to see point sources near the slit down 
to V$\sim$22. 

\subsection{Kitt Peak 4-m Telescope}

Observations were obtained with the RC spectrograph using grating KPC-22b in first (red) and second (blue) order. 
The blue spectral resolution for the setup used is
$\sim$5000, providing a wavelength coverage of 3700-5100\AA\ with a dispersion on the CCD detector of 0.72\AA/pixel. 
The red setup yielded 1.42\AA/pixel and covered 6200 to 6800\AA\~. 
The slit was set to 1 arcsec and used in an east-west (90 degree) orientation for all observations.
The weather was clear each night with stable 1.2 arcsec seeing. 
Observations of spectrophotometric standard stars were obtained near in time to the target
stars and used to provide relative flux values. FeAr, HeNeAr, or ThAr lamp exposures, obtained directly before or
after each spectrum, were used to set the wavelength scale.
Data reduction was done using IRAF 2-D and 1-D routines for spectroscopic data and produced
a final 1-D spectrum for each observation. 

\section{Discussion}

The stars are presented by their variable star name, for those which have one, otherwise we use their {\it Kepler}
input catalogue
(KIC) identification. The KIC in discussed in Brown et al. (2011) and can be accessed at the MAST
archive\footnote{http://archive.stsci.edu/kepler/kepler\_fov/search.php}. We note here that {\it Kepler} can
successfully perform photometric observations for sources near R$\sim$21-22 magnitude, with some dependence on the
variability amplitude of the source in question (e.g., deep eclipses or outbursts).

\subsection{V344 Lyr}

The {\it Kepler} light curve of V344 Lyr was presented in Still et al. (2010) and Wood et al. (2011)
where the star was shown to be a SU UMa type cataclysmic
variable with a photometrically determined orbital period of 2.11 hr. We report here on phase-resolved spectroscopy of 
the star obtained over the time period of 6:01 to 9:39 UT
on 2011 June 5. This time period covers 1.7 orbital cycles of V344 Lyr. 
Figure 1 shows one of the 12 time-resolved blue and red spectral pairs we obtained for
V344 Lyr. No significant differences are seen in any of the spectra or their 
emission line shapes 
over the 3 hr 40 min time span. The smooth and narrow emission line profiles seen in V344 Lyr suggest that
this binary has a low orbital inclination.

Measurements of the equivalent width of H$\alpha$ and H$\beta$ during the 1.7 orbital cycles
show nearly constant values of -32.5$\pm$2 \AA\ and -22$\pm$2 \AA\ respectively. 
Fitting a Gaussian profile to these emission lines in each spectrum, we found that 
the measured velocities show a low amplitude modulation (K$\sim$10 km/sec) consistent with a low
orbital inclination (Fig. 2). Higher order Balmer lines have a similar velocity appearance to that of 
H$\beta$. 
The velocities vary with a roughly 
2-hr period, more obvious in the H$\beta$ velocities, a time period 
consistent with the well measured photometric period for this star.
However, due to the narrow lines, their small K amplitude, our 
uncertainties of $\sim$6 km/sec per point, and long (20 min) integrations, a robust
sinusoidal (orbital) fit to the measurements was not possible.
While the photometric analysis presented in Wood et al. (2011) can
only limit the inclination of V344 Lyr to between 0 and 60 degrees, 
the small velocity variation observed most likely suggests that V344 Lyr 
has a very low inclination, near 5-10 degrees. 

\subsection{V358 Lyr}

The magnitude range ascribed to this star in DWS97
is $V$=16 to fainter than 20. 
Their associated finder chart shows only an empty circle at the star's position.
Antipin et al. (2004) provide a discussion of the star's history, including
its faint limit being below V=21 and some confusion as to the maximum magnitude reached during outburst. 
They conclude that V358 Lyr is a CV with rare and apparently large outbursts. 
DWS97, in agreement with Antipin's findings, list V358 Lyr as a nova or possibly 
a TOAD (Tremendous Outburst Amplitude Dwarf nova; Howell et al., 1995).
Our examination of the field using the Palomar 200" acquisition CCD camera revealed no 
source at the stars position down to 21.5-22 visual magnitude. 
This star had an outburst in 2008 (Kato et al. 2009) 
reaching a peak magnitude of 16.1 and revealing a superhump period
of 0.0556 days. Putting all of this information together, suggest that V358 Lyr has a short orbital
period and an outburst amplitude of $>$6 magnitudes, it is likely to be a TOAD.

\subsection{V452 Lyr}

V452 Lyr is a suspected dwarf nova with a minimum magnitude of V$>$18.5. We estimate that V452 Lyr 
was near $V$=20.5 when we obtained our spectrum
being nearly equal in brightness to its close neighbor sitting $\sim$5
arcsec to the southeast (see finding chart in DWS97). 

Our spectrum (Fig. 3), obtained presumably near minimum light,
shows a red source not a blue one. While of low S/N, the
blue spectrum appears almost featureless and without a rising blue continuum, the usual tell-take sign of an
accretion disk system. The red spectral region (bottom panel Fig. 3) is of low S/N and smoothed by 7
points in the plot.
The reddish spectral appearance of V452 Lyr is in agreement with 
the few available colors for this source obtained
from the MAST archive; g-R=0.8 mag and R-I=0.3i mag.
The {\it Kepler} light curve of V452 Lyr 
covers 263 days in long cadence (30 minute samples) and 67 days at short cadence (1 minute samples) 
and shows no statistically significant variations.
The spectral appearance and light curve behavior of the target we observed places doubt on the 
true identify of this variable star but it seems unlikely to be a CV.

\subsection{V585 Lyr}
This star has a minimum magnitude listed as fainter than $V$$\sim$21.5 in DWS97.
The 200" telescope CCD acquisition camera image revealed a very faint
source near the position of the star but with a magnitude estimated to be near 22. 
If the point source was V585 Lyr, it was too faint to get a spectrum
as a part of this survey program and thus we can not confirm the source or its CV nature.

\subsection{V516 Lyr}

This star is one of two cataclysmic variables 
in the {\it Kepler} field of view which may be a member of the open cluster NGC 
6791; V523 Lyr is the other (see below).
While a faint source, near 22, is shown on the DWS97 finder chart, our CCD acquisition image at Palomar
revealed no source present down to $\sim$22. Therefore, no spectrum was obtained and thus we can not
confirm the source or its CV nature. 

\subsection{V523 Lyr}

V523 Lyr is located on the outskirts of the core of the open cluster NGC 6791, 
thus its membership in that cluster
has long been suspected. Listed as a VY Scl or Z Cam star, this object has a historic magnitude
range quoted in DWS97
as 17.7 to 20.2.
Figure 4 presents the Quarter 6-8 long cadence {\it Kepler} light curve of V523 Lyr. 
The light curve of this star was normalized using the tools available at the NASA
Exoplanet archive and shows
clear indications of dwarf nova outbursts
as well as one superoutburst occurring near the end of the data set, starting about day 760.
The light curve covers $\sim$250 days and 
contains ten normal outbursts that occur almost periodically at $\sim$18 day intervals, each
lasting about 4 days. We see that the outbursts 
grow in amplitude as they approach the superoutburst, but in a non-linear step wise fashion.
The first normal outburst in the {\it Kepler} light curve, shown in the lower left of Figure 4, 
as well as the second outburst, appear to be two outbursts occurring
in close succession. We also note (bottom right of Figure 4) that directly after the superoutburst, 
some small amplitude, shorter duration ($\sim$1 day) outbursts occur. 
The entire set of {\it Kepler} long and short cadence light curves of V523 Lyr 
will be the subject of an upcoming detailed light curve study by our group. 

In between the outbursts, as easily seen in the top and bottom middle plots of Figure 4, is a periodic 
modulation in the light curve.
A period search, with the outbursts removed, identifies this period as 0.087478 days 
(2.1 hr), and we present a phase-binned light curve of V523 Lyr at minimum light
in Figure 5. 
Whether this period is the true orbital period of V523 Lyr
will be one of the findings of our
upcoming study. The light curve, showing a superoutburst,  and the (orbital) period of 
V523 Lyr suggest that it is a SU UMa type cataclysmic variable.

The spectrum of V523 Lyr is shown in Figure 6. The blue spectrum rises to the shorter wavelengths 
and shows H$\beta$ to be in weak emission while the
higher Balmer lines appear in absorption. 
The red spectrum shows a narrow H$\alpha$ emission line sitting on an otherwise featureless
continuum. These spectral features are not typical of a low state 
SU UMa star.  We note, however, that our spectrum was taken on JD 2455719,
two days into a normal outburst of V523 Lyr that started on day 885.
Thus, our Palomar spectrum was obtained during an outburst which would explain
its non-typical SU UMa minimum light appearance.

\subsection{V1504 Cyg}

Being a relatively bright CV, even at minimum light, V1504 Cyg has been studied in some detail with 
recent photometric papers being Pavlenko et al. (2002)
and Cannizzo et al. (2012). DWS97 lists V1504 Cyg as a SU UMa type CV ranging from $V=13.5$ to $17.4$ 
in magnitude and 
Cannizzo et al (2012)
provide a detailed study of the {\it Kepler} light curve of this star, confirming its period at 1.67 hr as well as its 
SU UMa status. 

Our Kitt Peak 4-m blue observation (Fig. 7) shows a classic dwarf nova spectrum with double peaked emission in the
higher Balmer series lines. He I (4471\AA) is also present in emission. The general spectral appearance, showing
double-peaked Balmer emission lines, 
argues for a moderate to relatively high orbital inclination (perhaps 40-60 degrees). 
We note that the spectrum presents a mostly flat Balmer decrement. 
This is a typical spectral appearance of a short period, low mass transfer CV. 

\subsection{KIC 9778689 = BOKS CV}

Feldmeier et al. (2010) discovered this variable blue source in a pre-launch survey of the {\it Kepler} field 
of view. They present a finding chart and a light curve suggesting a minimum magnitude 
near $V=20$ and an apparent dwarf nova type outburst rising
to $\sim$17th and lasting $\sim$5 days. Thus, KIC 9778689 was suspected of being a
U Gem type CV in the {\it Kepler} field. A CCD acquisition image taken at Palomar could not find a source at the target
position to magnitude 22, so no spectrum was obtained. The {\it Kepler} Quarter 6-8 
short cadence (1 minute sampling) 
light curve of this star (Figure 8) was normalized using the tools available at the NASA
exoplanet archive.
Covering 250 days, the light curve
is rather noisy due to the source faintness but shows no cataclysmic variable type 
outbursts of any kind nor
any variability similar to that reported by Feldmeier et al.
In fact, the light curve does not even appear to be variable except 
that it does show occasional low-level flaring, as seen in the bottom panels of Figure 8. 
This object may be transient source of some type, a TOAD, or not a CV at all. 

\subsection{KIC 11390659}

KIC 11390659 was observed with the KPNO 4m telescope and
RC Spectrograph on Sept 12, 2010 from 2:50-4:54 UT. 
On June 7, 2011, 11 additional spectra were obtained 
from 8:18 to 10:28 UT at
the Palomar 200" telescope using the Double Imaging Spectrograph.
In each case, the spectra were wavelength and flux calibrated
with IRAF routines and the line velocities were measured with the
``e" (centroid) and ``g" (Gaussian) routines within the splot package. A typical
spectrum from each run is shown in Figures 9 and 10.
These spectra are usual for a short period cataclysmic
variable, with strong Balmer emission lines and a steep
blue continuum. The Palomar data reveal a narrow peak
in the lines, which shifts back and forth, typical of
a hot spot on an accretion disk. In addition, the strong
blue continuum short-ward of 4000\AA\ is indicative of
the contribution of a hot white dwarf or boundary layer.

While the object is a known ROSAT source, 
(1RXSJ1858311.1+491434) with an X-ray
count rate of 0.19$\pm$0.02 c s$^{-1}$ obtained during
731 sec of observation, the absence of strong
HeII emission likely means it does not contain a highly
magnetic white dwarf. The {\it Kepler} public long cadence 
light curve (Figure 11), covers Quarters 6 and 7 (about 150 days), and shows
high amplitude variability on timescales of days and flare-like structures but no
periodicity was detected such as that which would be expected from an active accretion spot
on a polar or a white dwarf spin from an intermediate polar.

A least squares fit of the line velocities to a sine-curve
was used to determine $\gamma$ (systemic velocity),
K (semi-amplitude), and P (orbital period).The errors on
$\gamma$ and K as well as the standard deviation ($\sigma$)
of the fit to the data points were determined from a Monte Carlo method.
Due to the short length of the datasets, the periods are not well established
but the common solution for all 3 lines from the blue channel in
the Palomar data are in the range of 106-109 min and both datasets
are consistent with an orbital period in this range. Due to the
larger errors on the KPNO data, the period was fixed at 107 min
to provide a better constraint on the K amplitude. The radial 
velocity solutions are given in Table 3 and a plot of the
data and the sine fits for the Palomar data is shown in Figure 12.
The K amplitudes and line widths are all consistent with a
dwarf nova with a moderate inclination.

\subsection{KIC 3426313 = Blue 10}

KIC 3426313 is a newly discovered blue variable based on an optical spectroscopic search for UV bright 
sources. Using the Howell-Everett $UBV$ catalog of the {\it Kepler} field 
(available at the MAST archive; Everett, Howell, \& Kinimuchi 2012), 
and selecting sources with extremely blue $U-B$ colors ($U-B=-0.666 mag, B-V=0.356$ mag) 
this source was identified as a UV bright
object.
Based on its spectroscopic appearance (Figure 13), this star is likely a new cataclysmic variable in the 
{\it Kepler} field of view. The blue spectrum shows very narrow Balmer emission lines
along with broad absorption troughs in the continuum. The red spectrum shows weaker H$\alpha$ emission,
likely indicating emission line variability as the two spectra were obtained 4 months apart.
Further study will be required to confirm the true nature of this star.

\subsection{KIC 8490027 = Blue 19}

This star was selected in the same manner as described for KIC 3426313 and is also a
newly discovered cataclysmic variable candidate. KIC 8490027 appears to possibly be a
nova-like type (Figure 14) showing Balmer absorption profiles from a nearly
edge-on thick accretion disk or the underlying white dwarf. The red spectrum shows a weak H$\alpha$ line now not centered in 
its absorption trough, indicative of velocity motion differences between the two non-contemporaneous
spectra. The emission lines are narrow in this source perhaps
pointing to a low orbital inclination.

\section{Conclusion} 
We have provided information on eleven cataclysmic variables residing within the {\it Kepler} field 
of view. These stars have all been or soon will be observed by {\it Kepler} and their 
light curve observations are or will be public.
Seven of the stars were previously known. New spectroscopy is presented for seven sources
while no source could be confirmed for three of the stars. 
We presented the previously unpublished {\it Kepler} light curves for three of the
sources, one of which, V523 Lyr, is a possible member of the open cluster NGC 6791. 
Evidence provided by a spectrum or {\it Kepler} light curve observation calls
two candidates into question as being CVs; V452 Lyr and KIC 9778689 (BOKS CV).
Given that the {\it Kepler} mission has and will continue to provide nearly continuous photometric coverage for 
the majority of these stars, detailed phase-resolved spectroscopy will be 
important to allow their stellar and orbital parameters to be determined.
However, such study will require large telescopes for the systems that are fainter 
than $V$$\sim$20th magnitude.
Cataclysmic variables located within the {\it Kepler} field 
have a large potential to allow the first detailed view of the systems
as a whole. The photometric coverage of the stars is unprecedented and will not likely 
occur again for decades to come.
Studies of the accretion process, stellar variations, and details of short term and long term effects, aided by
spectroscopic work, can turn this set of stars into a Rosetta Stone for CV research.

We wish to thank the observatory staffs of Kitt Peak and Mount Palomar for their help in carrying out the observations
presented here. PS acknowledges support from NSF grant AST-1008734. MW acknowledges support from NSF grant AST-1109332.
{\it Kepler} was competitively selected as the 10th NASA Discovery mission.

{\it Facilities:} NOAO/Kitt Peak 4meter (RCSPEC), Palomar 200" (DBSP), {\it Kepler} Space Telescope


\clearpage

\begin{deluxetable}{cccccccl}
\tablenum{1}
\tablewidth{0in}
\tablecaption{Cataclysmic Variables in the {\it Kepler} Field (Historic and New)}
\tablehead{
\colhead{CV Name} & \colhead{KIC\tablenotemark{a}} & \colhead{RA} & \colhead{DEC} & \colhead{P$_{\rm{orb}}$ (hr)} & 
\colhead{$V$} & \colhead{Type} & \colhead{Notes} 
 }
\startdata
V344 Lyr & 7659570 & 18 44 39.17 & 43 22 28.2 & 2.11 & 13.8-20.0 & SU UMa     & 1,2 \\
         & 11390659 & 18 58 30.91 & 49 14 32.7 & 1.78 & 16.5 & CV  & 5 \\
V358 Lyr & \nodata & 18 59 32.95 & 42 24 12.2 & \nodata & 16.0-$>$21.5 & WZ Sge?  & DWS97 \\
V447 Lyr & 8415928 & 19 00 19.92 & 44 27 44.9 & 3.74 & 17.2-18.6 & U Gem  & b  \\
Blue10   & 3426313 & 19 03 38.36 & 38 32 14.9 & \nodata & 15.9 & DN?  & 5 \\
MV Lyr   & 8153411 & 19 07 16.29 & 44 01 07.7 & 3.18 & 12.1-17.7 & NL &  c  \\
V452 Lyr & 7742289 & 19 10 26.32 & 43 11 55.2 & \nodata & 17.6-18.5 & DN? & DWS93   \\
V585 Lyr & \nodata & 19 13 58.40 & 40 44 09.0 & \nodata & 14.9-21.1 & SU UMa? & DWS97 \\
Blue19   & 8490027 & 19 19 31.08 & 44 32 44.0 & \nodata & 16.7 & DN? & 5 \\
V516 Lyr & 2436450 & 19 20 35.73 & 37 44 52.3 & \nodata & 18.9-22.2 & DN? &  DWS97 \\
V523 Lyr & \nodata & 19 21 07.40 & 37 47 56.5 & 2.1   & 17.7-20.2 & SU UMa & 5 \\
V1504 Cyg & 7446357 & 19 28 56.47 & 43 05 37.1 & 1.67 & 13.5-17.4 & SU UMa &  3 \\
BOKSCV    & 9778689 & 19 40 16.19 & 46 32 48.0 & \nodata & 16.0-$>$22 & DN/NL? &  4,5 \\ 
\enddata 
\tablenotetext{a}{KIC = {\it Kepler} Input Catalog}
\tablenotetext{b}{Not discussed in this paper, see Ramsay et al. (2012)}
\tablenotetext{c}{Not discussed in this paper, see Linnell et al. (2005)}
\tablenotetext{1}{Wood et al. (2011)} 
\tablenotetext{2}{Still et al. (2010)}
\tablenotetext{3}{Cannizzo et al. (2012)}
\tablenotetext{4}{Feldmeier et al. (2010)} 
\tablenotetext{5}{This paper}
\end{deluxetable}

\begin{deluxetable}{ccccc}
\tablenum{2}
\tablewidth{0in}
\tablecaption{Spectroscopic Observing Log}
\tablehead{
\colhead{CV Name} & \colhead{KIC\tablenotemark{a}} & \colhead{Telescope} & \colhead{UT Date} & \colhead{Int Time (sec)}
 }
\startdata
V344 Lyr & 7659570 &  Palomar 200" & 5 Jun 2011 & 12 x 1200 \\
\nodata & 11390659  & Kitt Peak 4-m & 12 Sep 2010 & 11 x 600 \\
\nodata & 11390659  & Palomar 200" & 7 Jun 2011 & 11 x 720 \\
Blue10 (blue)    & 3426313 & Kitt Peak 4-m &  15 Jun 2012 & 900 \\
Blue10 (red)   & 3426313 & Kitt Peak 4-m &  8 Oct 2012 & 600 \\
V452 Lyr & 7742289 &  Palomar 200" & 7 Jun 2011 & 1200 \\
Blue19 (blue)   & 8490027 & Kitt Peak 4-m & 17 Jun 2012 & 1200 \\
Blue19 (red)   & 8490027 & Kitt Peak 4-m & 8 Oct 2012 & 900 \\
V523 Lyr & \nodata &  Palomar 200" & 7 Jun 2011 & 1800 \\
V1504 Cyg & 7446357 & Kitt Peak 4-m & 11 Sep 2010 & 720 \\
\enddata 
\tablenotetext{a}{KIC = {\it Kepler} Input Catalog}
\end{deluxetable}

\clearpage

\normalsize
\begin{deluxetable}{lccccc}
\tablenum{3}
\tablewidth{0pt}
\tablecaption{KIC 11390659 Radial Velocity Solutions}
\tablehead{
\colhead{Obs} & \colhead{Line} & \colhead{P (min)} & 
\colhead{$\gamma$ (km s$^{-1}$)\tablenotemark{a}} & \colhead{K (km s$^{-1}$)} &
\colhead{$\sigma$ (km s$^{-1}$)} }
\startdata
Palomar & H$\alpha$ & 103 & -37$\pm$2 & 82$\pm$9 & 19 \\
Palomar & H$\beta$ & 107 & -22$\pm$1 & 99$\pm$6 & 13 \\
KPNO & H$\beta$ & 107 (fixed) &-3$\pm$2 & 89$\pm$14 & 32 \\
Palomar & H$\gamma$ & 106 & -35$\pm$1 & 114$\pm$7 & 15 \\
KPNO & H$\gamma$ & 107 (fixed) & -43$\pm$1 & 90$\pm$9 & 23 \\
Palomar & H$\delta$ & 109 & -10$\pm$1 & 110$\pm$6 & 12 \\
\enddata
\tablenotetext{a}{No attempt was made to produce an absolute $\gamma$ velocity consistent across the different
telescopes and nights listed here.} 
\end{deluxetable}

\clearpage

\begin{figure}
\epsscale{0.75}
\includegraphics[angle=-90,scale=0.65,keepaspectratio=true]{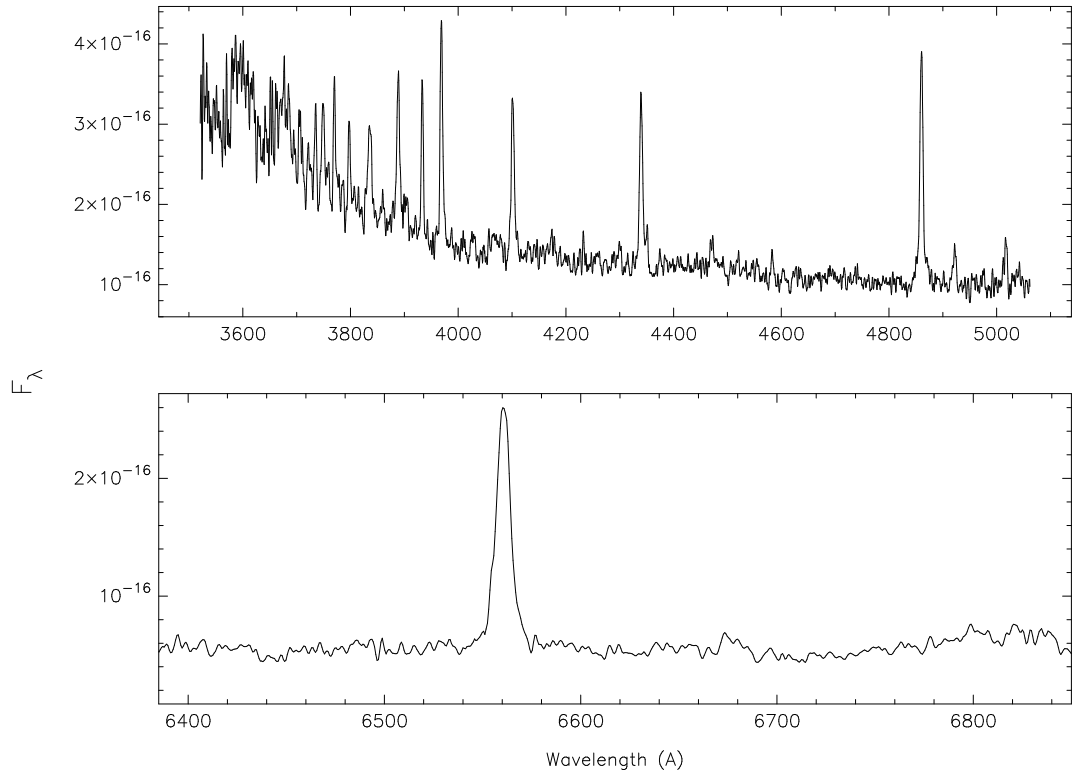}
\caption{Simultaneous red and blue spectra of V344 Lyr obtained with the DBSP at Mount Palomar. 
The blue spectrum shows both Balmer and He emission lines as well as a steeply rising blue continuum.
H$\alpha$ emission and weak He emission appear in the red spectrum.
The y-axis is flux in units of ergs/sec/cm$^2$/Angstrom.
}
\end{figure}

\begin{figure}
\epsscale{0.75}
\includegraphics[angle=-90,scale=0.65,keepaspectratio=true]{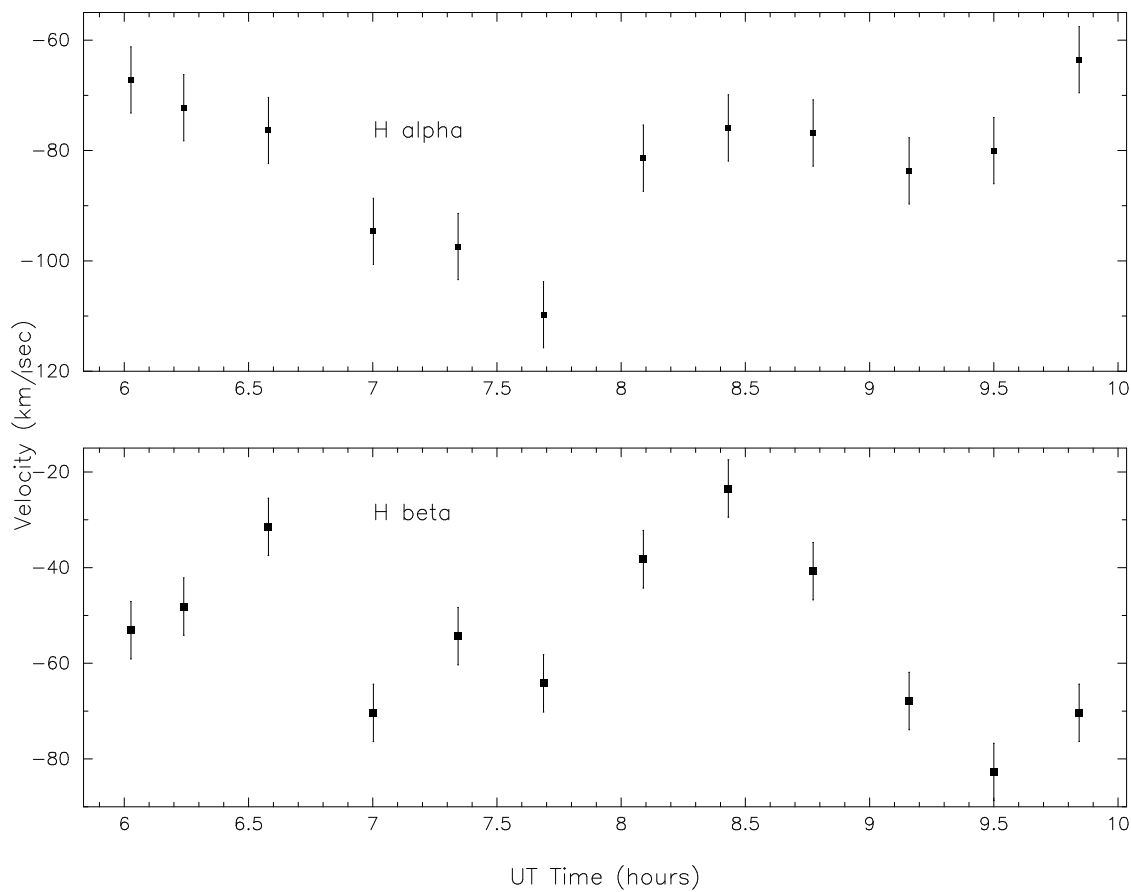}
\caption{Velocity measurements for H$\alpha$ (top) and H$\beta$ (bottom) for V344 Lyr. Each point
has a formal error of $\pm$6 km/sec. The time coverage is 1.7
orbital cycles and the data show a low amplitude modulation, K$\sim$10 km/sec, 
that appears to repeat approximately at 2
hours. This period is consistent with the well defined photometric period of 2.11 hours but due to the
low K amplitude, our spectral resolution, and the long (20 min) integrations, a robust RV fit was not
possible. 
}
\end{figure}
 
\begin{figure}
\epsscale{0.75}
\includegraphics[angle=-90,scale=0.65,keepaspectratio=true]{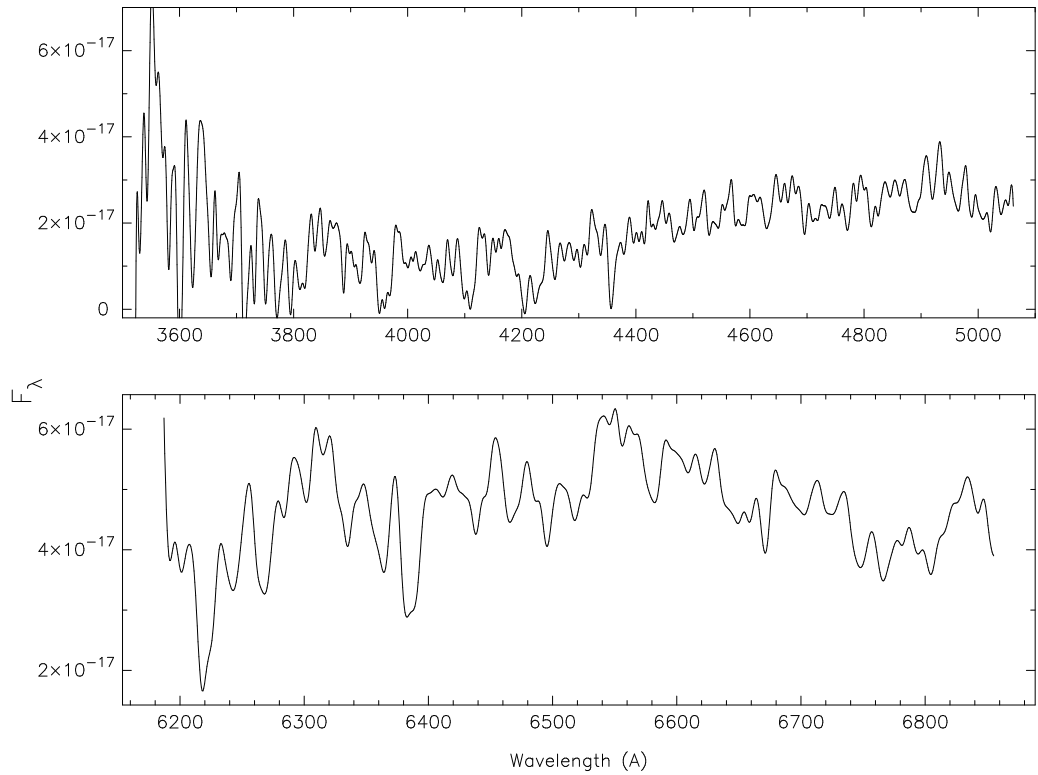}
\caption{Simultaneous red (smoothed by 7 points) and blue spectra of V452 Lyr. The general spectral appearance does not look like a typical
cataclysmic variable. The rising red continuum in the blue spectrum would be unusual for a CV. 
While of relatively low S/N, the spectra do not provide much evidence that V452 Lyr is indeed a CV.
The y-axis is flux in units of ergs/sec/cm$^2$/Angstrom.
}
\end{figure}

\begin{figure}
\epsscale{0.75}
\includegraphics[angle=-90,scale=0.65,keepaspectratio=true]{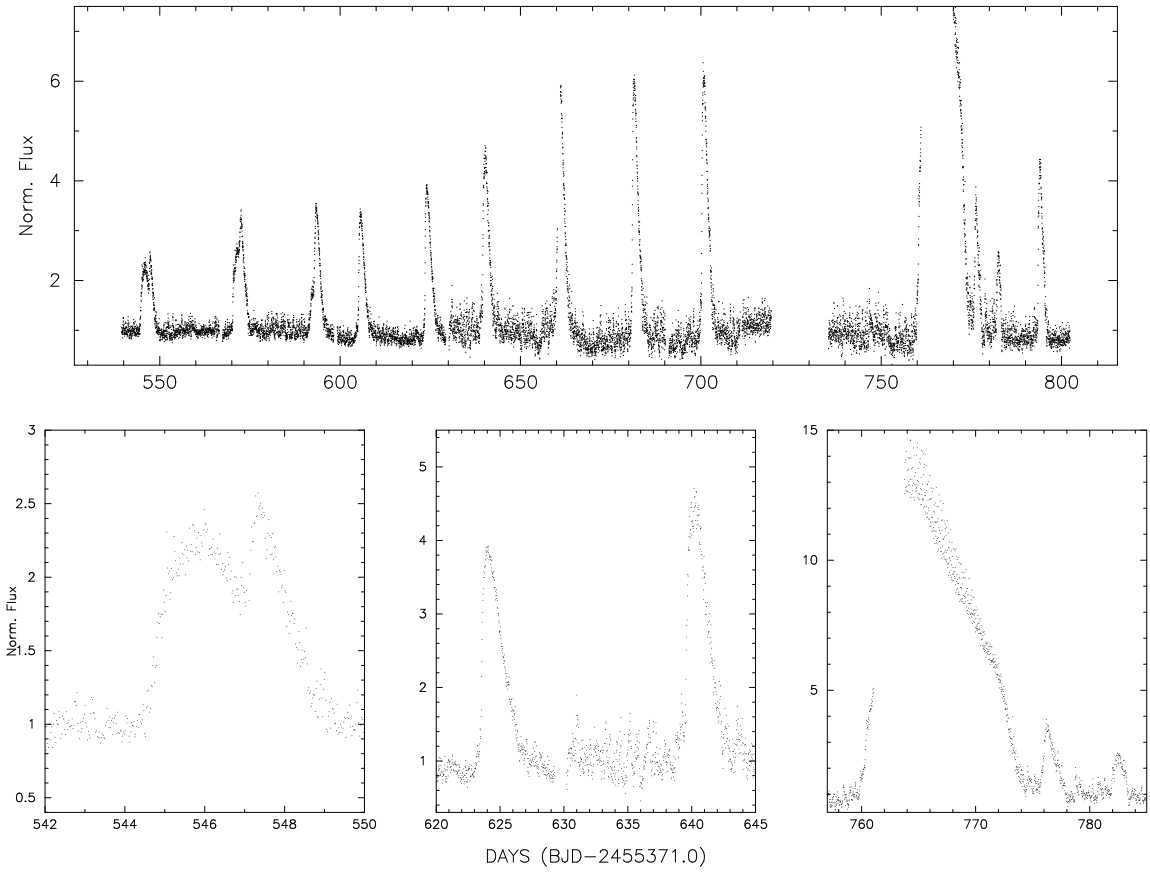}
\caption{{\it Kepler} long cadence (30 minute sample) light curve of V523 Lyr covering Quarters 6-8. 
The top plot shows the full light curve while
the bottom plots show detail at three interesting times. 
  }
\end{figure}

\begin{figure}
\includegraphics[angle=-90,scale=0.65,keepaspectratio=true]{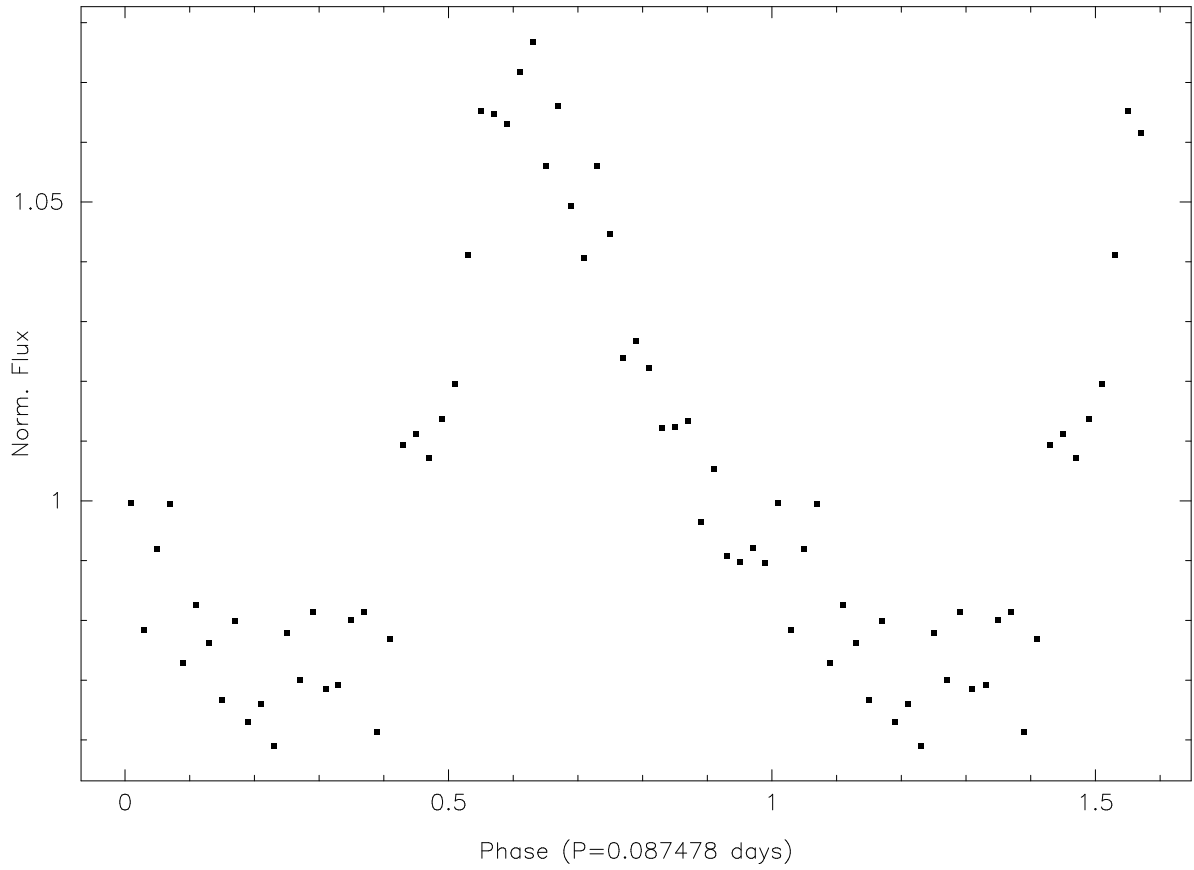}
\caption{Phased light curve of V523 Lyr using intra-outburst minimum light segments from the
data presented in the previous figure. The photometric period 
of 2.1 hr places this CV in the SU UMa class, confirmed by the superoutburst observed in the light curve.
  }
\end{figure}

\begin{figure}
\epsscale{0.75}
\includegraphics[angle=-90,scale=0.65,keepaspectratio=true]{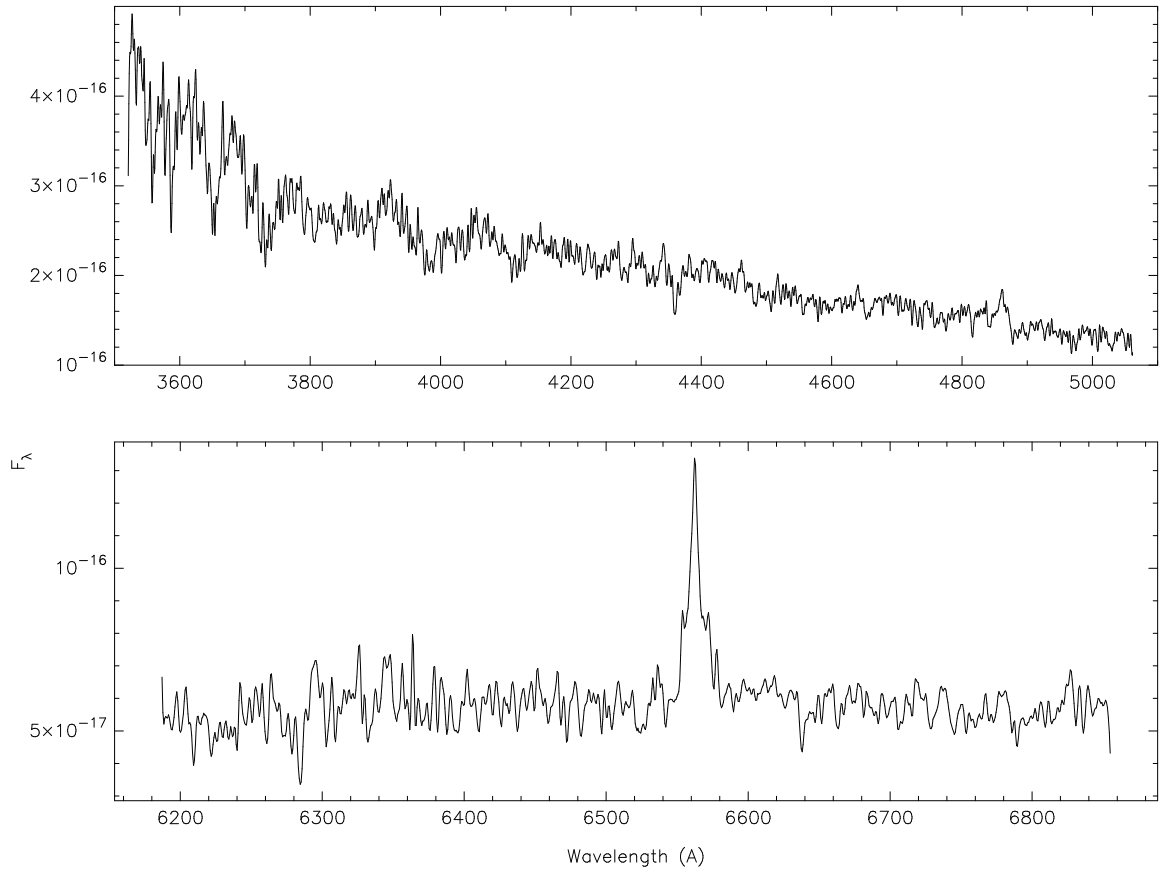}
\caption{Simultaneous red and blue spectra of V523 Lyr. H$\alpha$ and H$\beta$ are in emission while the rest of the
Balmer series is in absorption. This spectrum was obtained during an outburst (see text).
The y-axis is flux in units of ergs/sec/cm$^2$/Angstrom.
  }
\end{figure}
 
\clearpage

\begin{figure}
\includegraphics[angle=-90,scale=0.65,keepaspectratio=true]{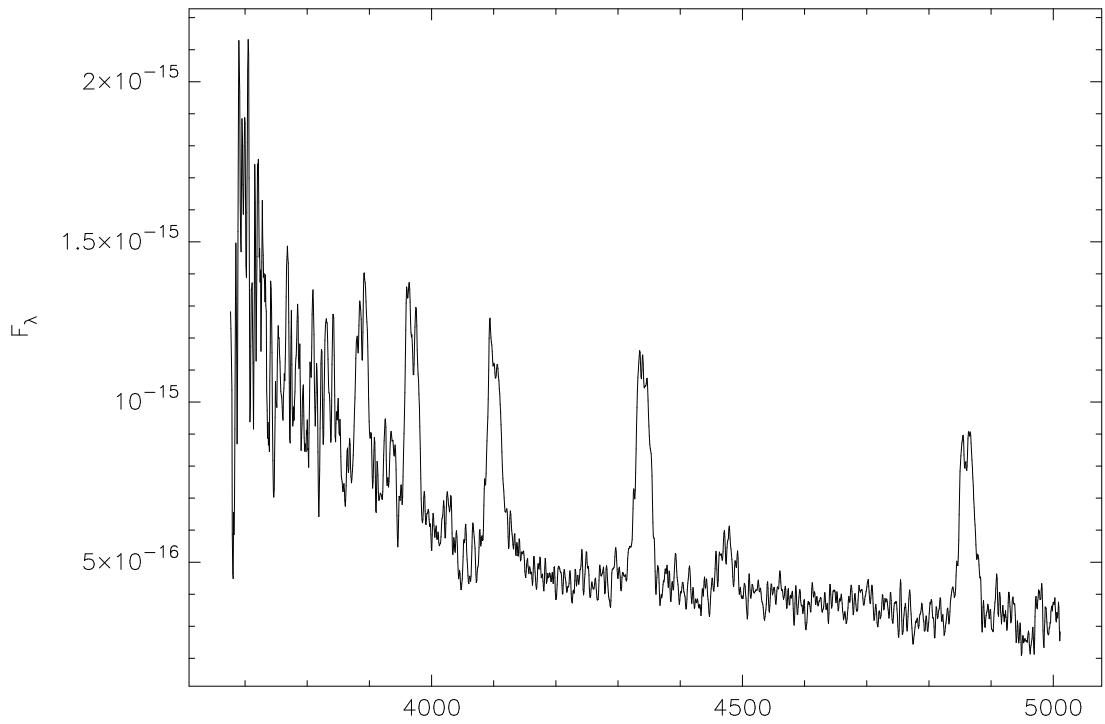}
\caption{Kitt Peak 4-m Blue spectrum of V1504 Cyg. The spectrum is a very classic high inclination cataclysmic variable
showing double-peaked Balmer emission lines from the accretion disk.
The y-axis is flux in units of ergs/sec/cm$^2$/Angstrom.
}
\end{figure}

\begin{figure}
\includegraphics[angle=-90,scale=0.65,keepaspectratio=true]{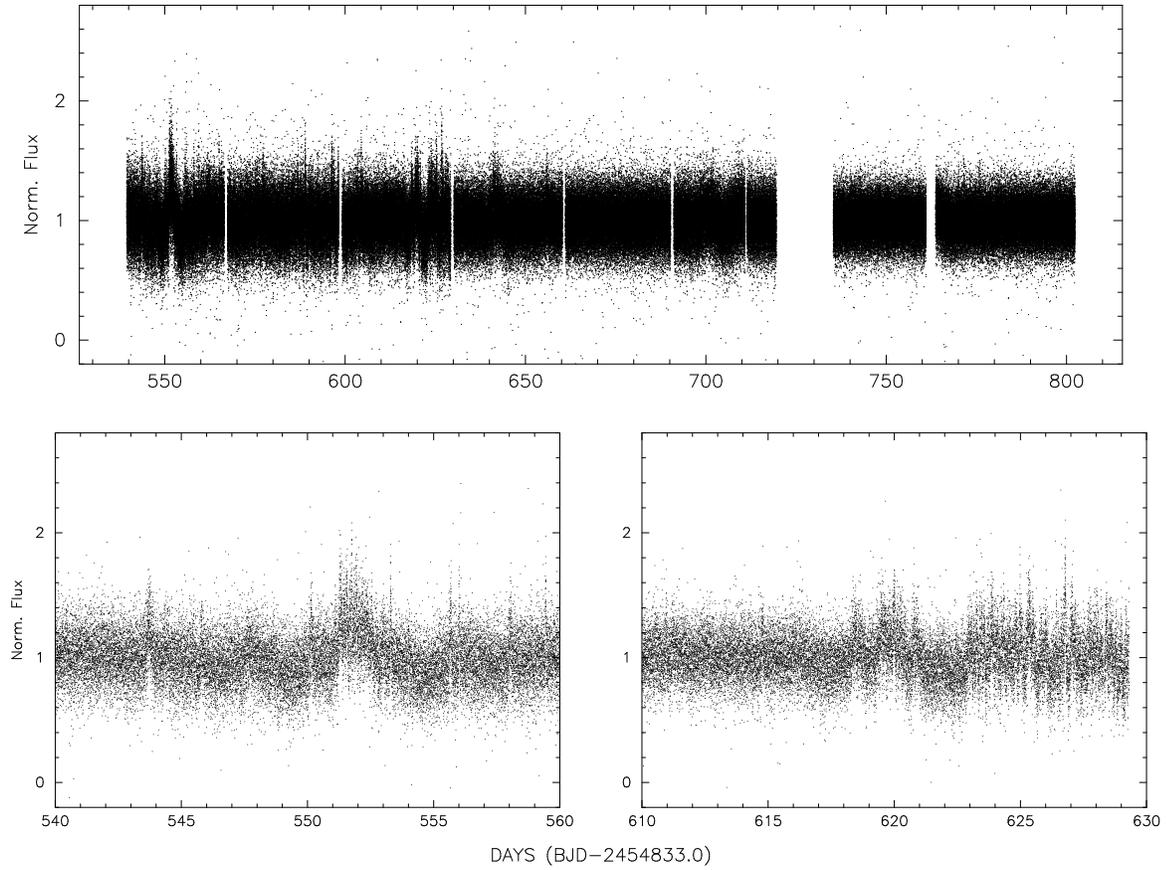}
\caption{{\it Kepler} short cadence (1 minute time samples) light curve of KIC 9778689 covering Quarters 6-8. 
The light curve shows essentially no
modulation except for occasional, small flare like events (see bottom zoomed plots). 
This source is not likely to be a cataclysmic variable.
}
\end{figure}

\begin{figure}
\plotone{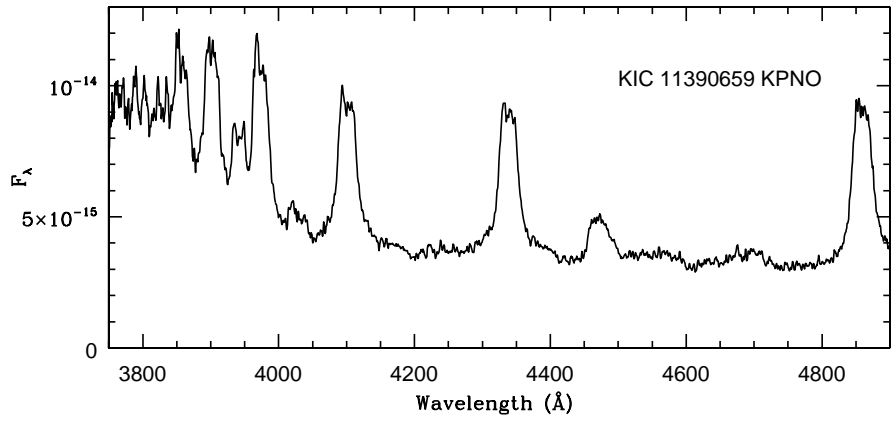}
\caption{Typical spectrum of KIC 11390659 obtained at the Kitt Peak 4-m telescope.
The y-axis is flux in units of ergs/sec/cm$^2$/Angstrom.
}
\end{figure}

\begin{figure}
\plotone{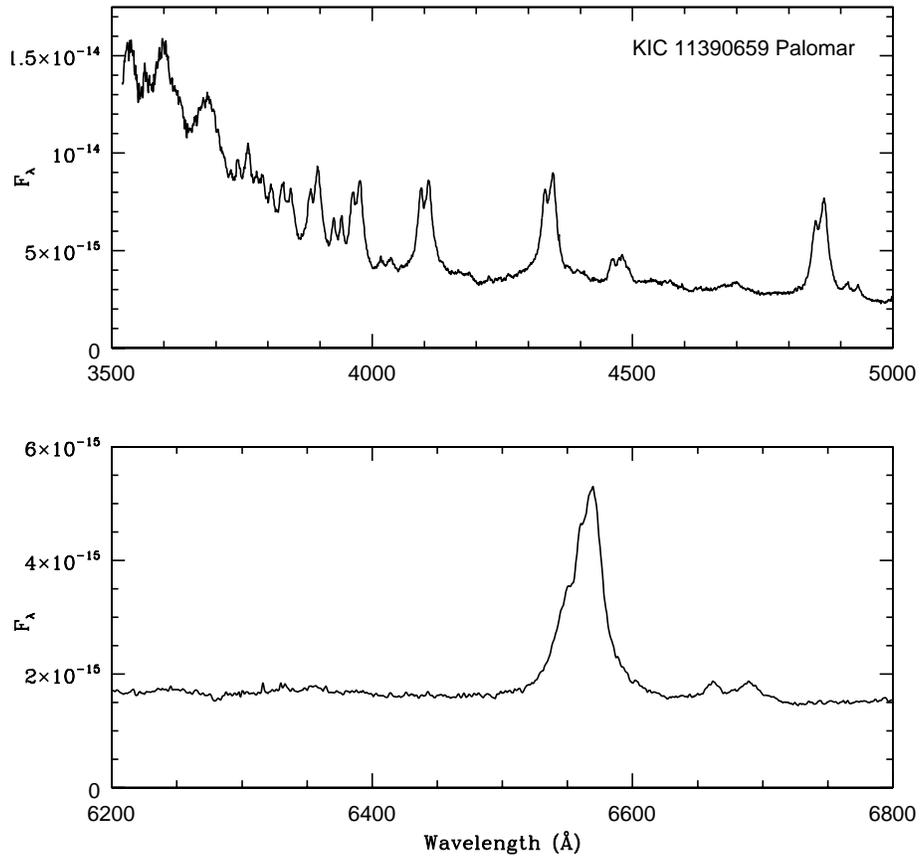}
\caption{Typical red and blue spectrum of KIC 11390659 obtained at the Mt. Palomar 200" telescope.
The y-axis is flux in units of ergs/sec/cm$^2$/Angstrom.
}
\end{figure}

\begin{figure}
\includegraphics[angle=-90,scale=0.65,keepaspectratio=true]{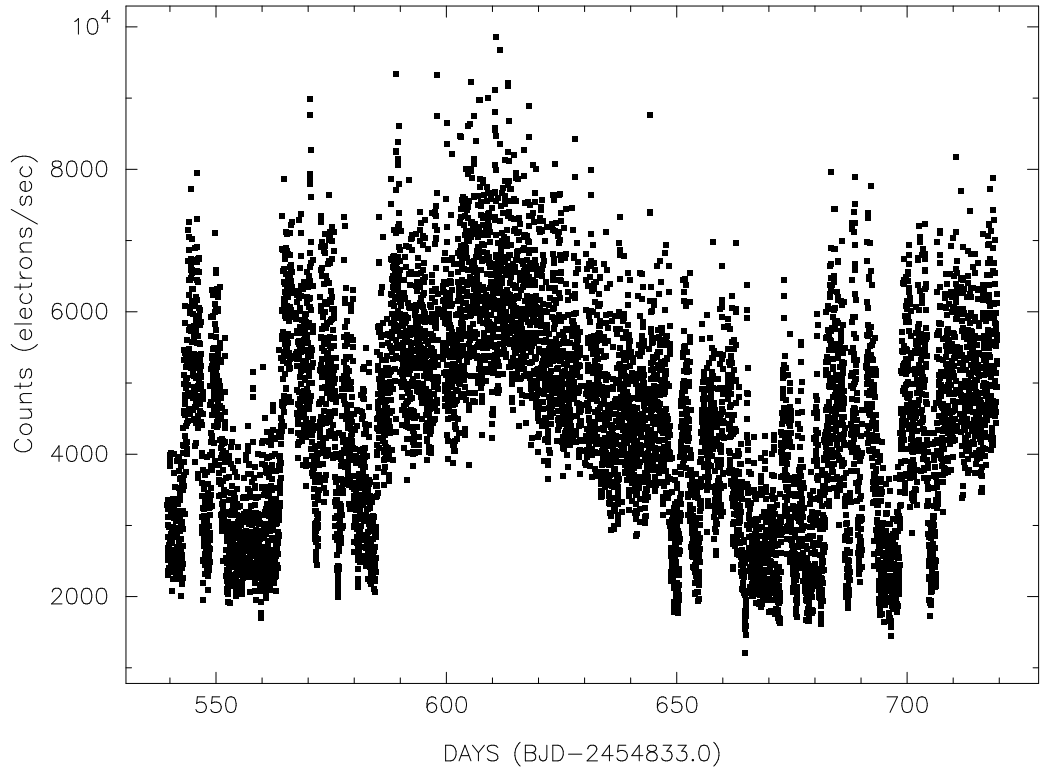}
\caption{{\it Kepler} light curve for KIC11390659 covering Quarters 6-7. 
The data show a rapid flare-like structure, each brightening lasting for a
few days, and a sine-like modulation possibly revealing a long term periodic structure. 
A power spectrum of this light curve
showed no significant periods including any near 107 minutes.
}
\end{figure}

\begin{figure}
\plotone{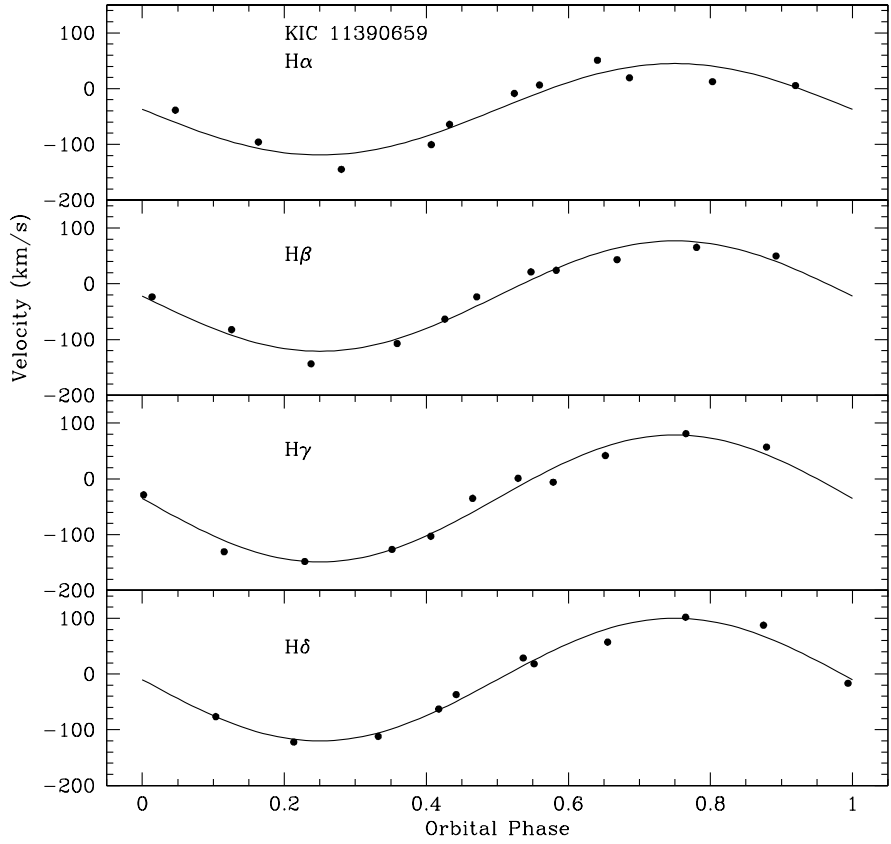}
\caption{Velocity curves for the four brightest Balmer lines in KIC 11390659. 
Each point has a formal 1-sigma velocity error of $\sim$6 km/sec. 
The best fit velocity curves 
as derived from the different emission line data give values for the 
period and phase which are consistent.
}
\end{figure}

\begin{figure}
\epsscale{0.75}
\includegraphics[angle=-90,scale=0.75,keepaspectratio=true]{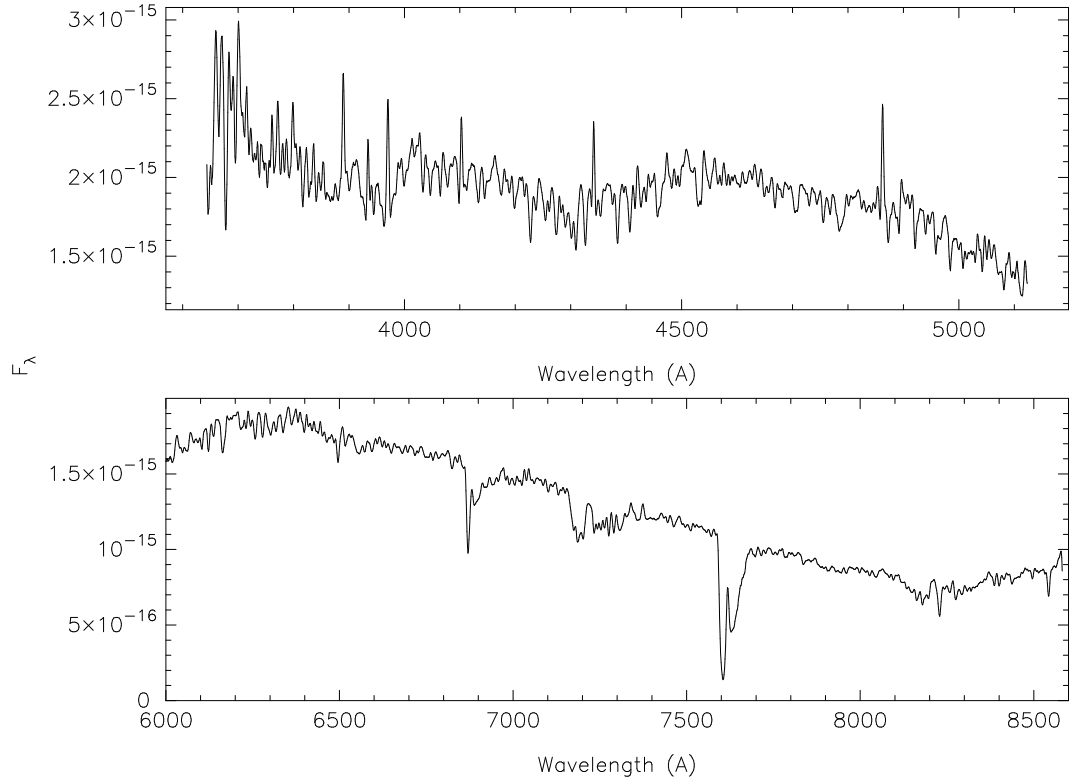}
\caption{Spectra of a newly discovered cataclysmic variable candidate. The top plot
shows a blue spectrum of KIC 3426313 (Blue 10) while the bottom covers the red spectral region for the star,
observed 4 months later.
KIC 3426313 shows very narrow Balmer emission in the blue with odd broad continuum features. 
The red spectrum shows a weak, narrow H$\alpha$ emission line.
The spectra are not corrected for telluric features
and the y-axis is flux in units of ergs/sec/cm$^2$/Angstrom.
}
\end{figure}

\begin{figure}
\epsscale{0.75}
\includegraphics[angle=-90,scale=0.75,keepaspectratio=true]{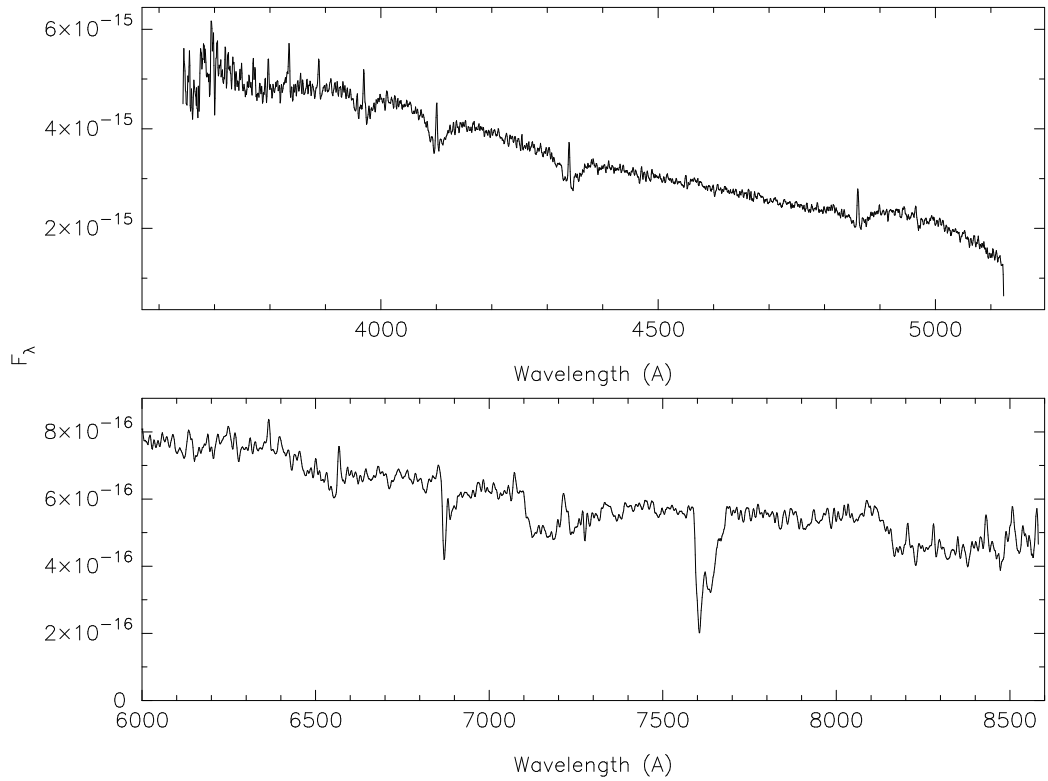}
\caption{Spectra of a newly discovered cataclysmic variable. The top plot
shows a blue spectrum of KIC 8490027 (Blue 19) while the bottom covers the red spectral region, observed 4 months later.
KIC 8490027 has narrow emission lines sitting within absorption features, possible from the accretion 
disk or the underlying white dwarf.
The spectra are not corrected for telluric features and
the y-axis is flux in units of ergs/sec/cm$^2$/Angstrom.
}
\end{figure}


\clearpage

\begin{thebibliography}{99}

\bibitem[]{} Antipin, S. V., Samus, N. N., and Kroll, P., 2004, IBVS, 5544
\bibitem[]{} Borucki, W. J., et al., 2010, Science, 327, 977
\bibitem[]{} Brown, T. M., Latham, D. W., Everett, M. E., \& Esquerdo, G. A., 2011, \aj, 142, 112
\bibitem[]{} Barclay, T., et al., 2012, \mnras, 422, 1219
\bibitem[]{} Cannizzo, J. K., et al., 2012, \apj, 747, 117
\bibitem[]{} Downes, R. A. \& Shara, M. M., 1993, \pasp, 105, 127
\bibitem[]{} Downes, R. A., Webbink, R., and Shara, M. M., \pasp, 1997, 109, 345
\bibitem[]{} Everett, M. E., Howell, S. B., and Kinimuchi, K., 2012, \pasp, 124, 316
\bibitem[]{} Feldmeier, J. J.,  et al., 2011, \aj, 142, 1
\bibitem[]{} Fontaine, G., et al., 2011, \apj, 726, 92 
\bibitem[]{} Howell, S. B., Szkody, P., and Cannizzo, J. K., 1995, \apj, 439, 337
\bibitem[]{} Kato, Y., et al., 2009, PASJ, 61, S395
\bibitem[]{} Linnell, A. P., et al., 2005, \apj, 624, 923
\bibitem[]{} Ostensen, R. H., et al., 2010, \mnras, 409, 1470
\bibitem[]{} Pavlenko, E. P., 2002, in ``The Physics of Cataclysmic Variables and Related Objects", 
ASP Conference Proceedings, Eds. B. T. Gänsicke, K.  Beuermann, \& K. Reinsch. Vol. 261, p523
\bibitem[]{} Ramsay, G., et al., 2012, \mnras, 425, 1479 
\bibitem[]{} Scaringi, S., Groot, P. J., Verbeek, K., Greiss, S., et al., 2012, \mnras, in press.
\bibitem[]{} Still. M., et al., 2010, \apj, 717, L113 
\bibitem[]{} Williams K. A.,  et al., 2010, \aj, 139, 2587
\bibitem[]{} Warner, B., 2003, ``Cataclysmic Variables", Cambridge University Press
\bibitem[]{} Wood, M. A., et al., 2011, \apj, 741, 105 

\end{thebibliography}
\end{document}